\documentclass{elsart}
\journal{Journal of Computational Physics}
\usepackage{amsmath,amssymb,graphicx}

\begin{document}

\begin{frontmatter}

\title{Absorbing boundary conditions for simulation of gravitational
waves with spectral methods in spherical coordinates}
\author{J\'er\^ome Novak}, \ead{Jerome.Novak@obspm.fr}
\author{Silvano Bonazzola} \ead{Silvano.Bonazzola@obspm.fr}

\address{Laboratoire de l'Univers et de ses Th\'eories (UMR8102 du
C.N.R.S.), Observatoire de Paris, Section de Meudon, F-92195 Meudon
Cedex France}

\date{November, 26th 2003} 

\maketitle

\begin{abstract}  
We present a new formulation of the multipolar expansion of an exact
boundary condition for the wave equation, which is truncated at the
quadrupolar order. Using an auxiliary function, that is the 
solution of a wave equation on the sphere defining the outer boundary
of the numerical grid, the absorbing boundary condition is simply
written as a perturbation of the usual Sommerfeld radiation boundary
condition. It is very easily implemented using spectral methods in
spherical coordinates.
Numerical tests of the method show that very good accuracy 
can be achieved and that this boundary condition has the same
efficiency for dipolar and quadrupolar waves as the usual Sommerfeld boundary
condition for monopolar ones. This is of particular
importance for the simulation of gravitational waves, which have
dominant quadrupolar terms, in General Relativity.
\end{abstract}
\end{frontmatter}


\section{Introduction}

\subsection{Wave equations in General Relativity}

The determination of numerical solutions of the Einstein equations is
the scope of {\em 
numerical relativity}. It is a fundamental issue not only for the
determination of gravitational wave signals for detector data
analysis, but also for the study of the properties of relativistic
astrophysical objects \cite{cm1}. Within numerical relativity studies,
the most commonly used formulation of the Einstein equations is the
so-called ``3+1'' formalism (also called {\em Cauchy formalism}
\cite{Yor79}) in which space-time is foliated by a family of space-like
hypersurfaces $\Sigma_t$, which are described by their 3-metric
$\gamma_{ij}$. The 4-metric $g_{\mu\nu}$ is then described in terms of
$\gamma_{ij}$, a 3-vector $N^i$ (called {\em 
shift}) and a scalar $N$ (called {\em lapse}). In this formalism,
the Einstein equations can be decomposed into a set of four constraint
equations and six second-order dynamical
equations. Solving the Einstein equations then turns to
be a Cauchy problem of evolution under constraints and there remains
the freedom to choose the time coordinate (slicing) and the spatial gauge.

For example, the choice of {\em maximal slicing} for the time coordinate (see
\cite{SY78}) converts the constraint equations to scalar form and a vectorial
Poisson-like equation, for which a numerical method for solution has
been presented in \cite{Phil}. As far as evolution equations are concerned,
they consist of six non-linear scalar wave equations in curved
space-time, with the additional choice of the {\em Dirac} gauge \cite{SY78}.
The whole system is a mixed initial value-boundary problem, and this
paper deals with boundary conditions for the time evolution
equations. Indeed, a simpler 
problem is considered: the initial value-boundary problem for a linear
and flat scalar wave equation:
\begin{equation}
\Box \phi(t,r,\theta,\varphi) = \sigma(t,r,\theta,\varphi) \label{e:defonde}
\end{equation}
where
$$\Box \phi = \frac{\partial^2 \phi}{\partial t^2} - \frac{\partial^2
\phi}{\partial r^2} - \frac{2}{r}\frac{\partial \phi}{\partial r} 
-\frac{1}{r^2}\left( \frac{\partial^2 \phi}{\partial \theta^2} +
\frac{1}{\tan \theta}\frac{\partial \phi}{\partial \theta} +
\frac{1}{\sin^2 \theta} \frac{\partial^2 \phi}{\partial \varphi^2}
\right)$$ is the usual flat scalar d'Alembert operator in spherical
coordinates $(r,\theta,\varphi)$ and $\sigma$ is a source. To solve a
more general problem in curved space-time, like for example:
\begin{equation}
\frac{\partial^2 \phi}{\partial t^2} - \mu^2(t,r) \Delta \phi = \phi^2,
 \label{e:ondecourbe} 
\end{equation}
one can put non-linear terms to the source $\sigma$ and represent at
each time-step the metric function $\mu^2$ by a polynomial
(semi-implicit scheme, see
\cite{Nov98} for an example in spherical symmetry).

\subsection{Motivations for general quadrupolar absorbing boundary conditions}

The study of the simple wave equation and its properties concerning
quadrupolar waves is more than a toy-model for numerical
relativity. There are many degrees of freedom in the formulation of
the Einstein equations and in the gauge choice. It is not clear which
of these formulations are well-posed or numerically stable
\cite{FriG00}. It is therefore important to have numerical tools
that are general in the sense that they can be used within the
framework of various formulations and gauges. Still, in many cases, the
dynamical degrees of freedom of the gravitational field can be
described by wave-like propagation equations in curved space-time. On
the other hand, since we are mainly interested in the gravitational
wave signal, which has a quadrupolar dominant term, we have to make
high precision numerical models (including boundary conditions) to
study this mode, as well as lower multipoles.

These statements can be illustrated as follows. One of the main sources
we want to study are binaries of two compact objects (neutron star or
black hole) orbiting around each other. Gravitational
waves  take away angular momentum and the system coalesces. In some
perturbative approach, the terms corresponding to this
``braking force'' result from a subtle cancellation between terms of
much higher amplitude \cite{Bla00}. In numerical non-perturbative
studies, the same phenomenon may happen and, if the dominant modes of
the wave
are not computed with enough precision, the angular momentum loss may
be strongly overestimated. Moreover, the time-scale for coalescence is
much larger than the orbital period and the system is almost
stationary. 

There has been many interesting developments concerning absorbing
boundaries in the last years, with the Perfectly Matched Layers (PML,
see \cite{Hes98} and \cite{LioMV02}) which consist in surrounding the
true domain of interest by an absorbing layer where the wave is
damped. These methods may not be the best-suited for our problems
since, as stated above, we might have to change the formulation of the
equations we want to solve. Moreover, the main problem we want to
address is the simulation of quadrupolar waves and, as it will be
shown later in this paper, with our formulation it is possible to have
a clear control on the behavior of these quadrupolar waves. Finally, this
formulation is straightforward to implement and very little CPU time
consuming in the context of spectral methods and spherical
coordinates, which we are already using to solve elliptic partial
differential equations (PDE) arising in numerical relativity (scalar
and vectorial ones, see \cite{Phil}). The development and
implementation of the PML techniques for our problem would require
much more work and computing time, whereas it is not guaranteed at all
it would give better results. For all these reasons we chose to
develop a new formulation of the Bayliss and Turkel \cite{BayT80}
boundary conditions, particularly well suited for using with spectral
methods and spherical coordinates.

The paper deals with this new formulation as well as numerical tests. It
is organized as follows. First, Sec.~\ref{s:BC} presents boundary
conditions: it briefly recalls main results from Bayliss and
Turkel (\ref{ss:mpoles}) and we then derive the formulation adapted
up to quadrupolar modes of the wave (\ref{ss:sphwave}). Then,
Sec.~\ref{s:tests} briefly describes spectral methods in spherical
coordinates that were used (\ref{ss:specmeu}) and details the numerical
results (\ref{ss:sorti}). Finally, Sec.~\ref{s:conc} gives a summary
and some concluding remarks.

\section{Absorbing boundary conditions}\label{s:BC}

An important difference between the solution of the wave equation and that
of the Poisson equation (as in \cite{Phil}) is the fact that
boundary conditions cannot be imposed at infinity, since one cannot
use ``compactification'', i.e. a change of variable of the type
$u=1/r$. This type of compactification is not compatible with
an hyperbolic PDE, see
\cite{Som49}. One has to construct an artificial boundary and
impose conditions on this surface to simulate an infinite
domain. These conditions should therefore give no reflection of the
wave, that could spuriously act on the evolution of the
system  studied inside the numerical grid. The boundary
conditions have to 
{\em absorb} all the waves that are coming to the outer limit of the
grid. The general {\em condition of radiation} is derived e.g. in
\cite{Som49}, and defined as
\begin{equation}
\lim_{r\to \infty} \left( \frac{\partial }{\partial r} +
\frac{\partial }{\partial t} \right) (r \phi)= 0. \label{e:somer0}
\end{equation}
At a finite distance $r=R$ the condition, which is then approximate,
reads 
\begin{equation}
\left. \frac{\partial \phi}{\partial t} + \frac{\phi}{R} +
\frac{\partial \phi}{\partial r} \right|_{r=R} = 0, \label{e:somerf}
\end{equation}
which will be hereafter referred as the ``Sommerfeld condition'' and
is exact only for pure monopolar waves. A completely general and exact
boundary condition for the wave equation on an artificial spherical
boundary has recently been derived by Aladl {\it et al.\/}
\cite{AlaDR02} and involves an 
infinite series of inverse Fourier transforms of the solution.
This condition may not be suitable for direct numerical implementation
for which Aladl {\it et al.\/} derived a truncated approximate condition.

\subsection{Asymptotic expansion in terms of multipolar
momenta}\label{ss:mpoles} 

A rather general method to impose non-reflecting boundary conditions is
to construct a sequence of boundary
conditions that, for each new term, are in some sense giving better
results. Some of the 
possibilities to define ``better'' are when the reflected wave decreases:
\begin{itemize}
\item as the incident wave approaches in a direction closer to some
preferred direction(s) (see e.g. \cite{EngM77}),
\item for shorter wavelengths,
\item as the position of artificial boundary goes to infinity.
\end{itemize}
This last approach is the most relevant to the problem of solving
the Einstein equation for isolated systems. It is also a way of 
expanding condition (\ref{e:somer0}) in terms of asymptotic series,
which  has been studied in \cite{BayT80}, where a sequence of
recursive boundary conditions is derived. Let us recall here some of their
results.

A radiating solution of (\ref{e:defonde}) with the source $\sigma = 0$
can be written as the following expansion:
\begin{equation}
\phi(t,r,\theta,\varphi) = \sum_{k=1}^\infty
\frac{f_k(t-r,\theta,\varphi)}{r^k}. \label{e:expand}
\end{equation}
The operators acting on a
function $f(t,r,\theta, \varphi)$ are recursively defined by: 
\begin{eqnarray}
B_1 f &=& \frac{\partial f}{\partial t} + \frac{\partial f}{\partial r}
+ \frac{f}{r}, \label{e:B1}\\
B_{n+1} f &=& \left( \frac{\partial}{\partial t} + \frac{\partial
}{\partial r}  + \frac{2n+1}{r} \right)B_n f . \label{e:Bn}
\end{eqnarray}
The family of boundary conditions then reads:
\begin{equation}
\left. B_n \phi\right|_{r=R} = 0. \label{e:BC+}
\end{equation}
In \cite{BayT80}, it is shown that, following from (\ref{e:expand}), a
radiating solution of the wave equation verifies:
\begin{equation}
B_n \phi = O\left( \frac{1}{r^{2n+1}} \right), \label{e:asymp}
\end{equation}
which in particular means that condition (\ref{e:BC+}) is an
asymptotic one in powers of $1/r$. The condition $B_1 \phi =0$ is 
same as the Sommerfeld condition (\ref{e:somerf}) and the same as the first
approximation in terms of the angle between the direction of
propagation of the wave and the normal to the boundary, derived in
\cite{EngM77}.

Finally, using expression (\ref{e:expand}) one can verify that the
operator $B_n$ annihilates the first $n$ terms of the
expansion. Thinking in terms of spherical harmonics, this means that
condition (\ref{e:BC+}) is exact if the wave carries only terms
with $l\leq n-1$. In other words, the reflection coefficients for all
modes lower than $n$ are zero. Since we are interested in the study of
gravitational wave emission by isolated systems, it is of great
importance to have a very accurate description of the quadrupolar part
of the waves, which is dominant. Therefore, if the $l=2$ part
of the gravitational wave 
is well described, higher-order terms may not play such an important
role in the dynamical evolution of the system. The situation then is
not so bad even if only an approximate boundary condition is imposed for
those terms with $l\geq 3$. Moreover, the error on the function
scales like $1/R^{n+1}$ so, if we impose 
\begin{equation}
\left. B_3 \phi \right|_{r=R} = 0, \label{e:B3}
\end{equation}
we have an exact boundary condition
for the main contribution to the 
gravitational wave and an error going to zero as $O(1/R^4)$. When
developing this expression, one gets:
\begin{eqnarray}
\left( \frac{\partial^3}{\partial t^3} +
3\frac{\partial^3}{\partial t^2 \partial r} + 
9\frac{1}{r}\frac{\partial^2}{\partial t^2} +
3\frac{\partial^3}{\partial t \partial r^2} + 
18\frac{1}{r^2}\frac{\partial}{\partial t} + 
18\frac{1}{r}\frac{\partial^2}{\partial t \partial r} +
\right. \nonumber \\
\left. \left. \frac{\partial^3}{\partial r^3} +
9\frac{1}{r}\frac{\partial^2}{\partial r^2} +
18\frac{1}{r^2}\frac{\partial}{\partial r} + 
6\frac{1}{r^3}
\right) \phi \right|_{r=R} = 0 . \label{e:B3devel}
\end{eqnarray}

\subsection{New formulation for quadrupolar terms}\label{ss:sphwave}

Starting from (\ref{e:B3devel}) and considering that $\phi$ is a
solution of the wave equation (\ref{e:defonde}), we replace second
radial derivatives with:
\begin{equation}
\frac{\partial^2 \phi}{\partial r^2} = \frac{\partial^2 \phi}{\partial
t^2} - \frac{2}{r}\frac{\partial \phi}{\partial r} -
\frac{1}{r^2}\Delta_{\text{ang}} \phi, \label{e:remplace}
\end{equation}
where:
\begin{equation}
\Delta_{\text{ang}} \phi = \frac{\partial^2 \phi}{\partial \theta^2} +
\frac{1}{\tan \theta}\frac{\partial \phi}{\partial \theta} +
\frac{1}{\sin^2 \theta} \frac{\partial^2 \phi}{\partial \varphi^2}
\end{equation}
is the angular part of the Laplace operator. We are making here the
assumption that, at the outer boundary of the grid ($r=R$), the source
term $\sigma$ of (\ref{e:defonde}) is negligible. This is a very good
approximation for our studies of isolated systems and is also the
assumption made when writing a solution to the wave equation in the
form (\ref{e:expand}). For example, the third order radial derivative
is replaced with
\begin{equation}
 \frac{\partial^3 \phi}{\partial r^3} = \frac{\partial^3
 \phi}{\partial t^2 \partial r} + \frac{2}{r^3} \Delta_{\text{ang}}
 \phi - \frac{1}{r^2}\Delta_{\text{ang}} \frac{\partial \phi}{\partial r}
+ \frac{2}{r^2}\frac{\partial \phi}{\partial r} -
 \frac{2}{r}\frac{\partial^2 \phi}{\partial r^2};
\end{equation}
and the second-order radial derivatives of the last term (combined with its
counterpart term in (\ref{e:B3devel}) ) is replaced once more using
(\ref{e:remplace}). The boundary condition is then written as:
\begin{eqnarray}
B_3 \phi = \left( 4 \frac{\partial^3}{\partial t^3} +
4\frac{\partial^3}{\partial t^2 \partial r} + 
16\frac{1}{r}\frac{\partial^2}{\partial t^2} +
18\frac{1}{r^2}\frac{\partial}{\partial t} + 
12\frac{1}{r}\frac{\partial^2}{\partial t \partial r} +
6\frac{1}{r^2}\frac{\partial}{\partial r} + 
\right. \nonumber \\
\left. -\frac{3}{r^2}\Delta_{\text{ang}} \frac{\partial }{\partial t} 
- \frac{1}{r^2}\Delta_{\text{ang}} \frac{\partial }{\partial r}
- \frac{5}{r^3} \Delta_{\text{ang}}
 + \frac{6}{r^3}
\right) \phi  . \label{e:B3modif}
\end{eqnarray}

We use the auxiliary function $\xi$:
\begin{equation}
\forall (t,\theta, \varphi), \quad \left. B_1\phi \right|_{r=R} =
\left. \left( \frac{\partial}{\partial t} 
+ \frac{\partial}{\partial r} + \frac{1}{r} \right) \phi(t,r,\theta,
\varphi) \right|_{r=R} = \xi(t,\theta, \varphi), \label{e:defxi}
\end{equation}
which is defined on the sphere at $r=R$. Inserting this definition into
the boundary condition $\left. B_3 \phi \right|_{r=R} = 0$, with
Eq.~(\ref{e:B3modif}), one gets:
\begin{equation}
\frac{\partial^2 \xi}{\partial t^2} -
\frac{3}{4R^2}\Delta_{\text{ang}}\xi + \frac{3}{R}\frac{\partial
\xi}{\partial t} + 
\frac{3\xi}{2R^2}
= \frac{1}{2R^2}\Delta_{\text{ang}} \left( \frac{\phi}{R} -
\left. \frac{\partial \phi}{\partial r} \right|_{r=R}
\right); \label{e:ondsph} 
\end{equation}
which is a wave-like equation on the outer boundary of the grid, with
some source term, equal to zero if the solution $\phi$ is spherically
symmetric. The boundary condition (\ref{e:B3}) is now equivalent to
the system (\ref{e:defxi})-(\ref{e:ondsph}). Written in this way, this
formulation can be
regarded as a perturbation of the Sommerfeld boundary condition ($B_1
\phi = 0$) given by (\ref{e:defxi}). The main advantages are that it
can be very easily implemented using spectral methods and spherical
coordinates (see Sec.~\ref{ss:specmeu}) and that mixed derivatives
have almost disappeared: there is only one remaining as a source of
(\ref{e:ondsph}).

\section{Numerical experiments}\label{s:tests}

\subsection{Implementation using spectral methods and spherical
coordinates}\label{ss:specmeu} 

Spectral methods (\cite{GotO77}, \cite{CanHQZ88}, for a review see
\cite{Boy01}) are a very powerful approach for the solution of a PDE
and, in particular, they are able to represent functions and their spatial
derivatives with very high accuracy. As presented in \cite{BonGM99},
we decompose scalar fields on spherical harmonics  $Y_l^m(\theta,
\varphi)$, for the angular part: 
\begin{equation}
\phi(t,r,\theta,\varphi) = \sum_{l=0}^L \sum_{m=-l}^l \phi_{lm}(t,r)
Y_l^m(\theta, \varphi),
\end{equation}
and on even Chebyshev polynomials $\left( T_{2k}(x=r/R) \right)$ for
the radial part of each $\phi_{lm}(t,r)$. Time derivatives are
evaluated using finite-difference methods. Since Chebyshev collocation
points are spaced by a distance of order $1/N^2$, (where $N$ is the
highest degree of the Chebyshev polynomials used for the radial
decomposition) near grid boundaries, the Courant condition on the time
step for explicit integration schemes of the wave equation
(\ref{e:defonde}) also varies like $1/N^2$. This condition is very
restrictive and it is therefore necessary to use an implicit
scheme. We use the Crank-Nicholson scheme, which is unconditionally
stable, as shown by various authors (see e.g. \cite{GotO77}). This
scheme is second-order in time and the smoothing of the solution due
to implicit time-stepping remains lower than the other errors
discussed hereafter. This implicit scheme results in a boundary-value
problem for $\phi$ at each time-step.  The solution to this problem is
obtained by inverting the resulting spatial operator acting on $\phi$
using the tau method. Its matrix (in Chebyshev coefficient space) has
a condition number that is rapidly increasing with $N$. This can be
alleviated by the use of preconditioning matrices, obtained from
finite-differences operators (see \cite{CanHQZ88}).

At the beginning of time integration, we suppose that $\phi$
satisfies the Sommerfeld boundary condition (\ref{e:somerf}), that is
$\forall (\theta, \varphi)\ \xi(t=0, \theta, \varphi) = 0$. $\xi$ is
then calculated at next time-step using 
(\ref{e:ondsph}). This is done very easily since the angular parts of
$\phi$ and $\xi$ are decomposed on the basis of spherical harmonics; each
component $\xi_{lm}(t)$ is the solution of a simple ODE in time, which
is integrated using the same Crank-Nicholson scheme as for the main
wave equation (\ref{e:defonde}), with boundary
conditions such that $\xi$ is periodic on the sphere. This is
already verified by the $Y_l^m$ (Galerkin method). We get, with
$\delta t$ being the time-step, $\phi^J_{lm}(r) =
\phi_{lm}(t+J\delta t, r)$ and $\xi_{lm}^J = \xi_{lm}(J\delta t)$:
\begin{eqnarray*}
\frac{\xi_{lm}^{J+1} - 2\xi_{lm}^J + \xi_{lm}^{J-1}}{\delta t^2}&+&
\frac{3}{8} \frac{l(l+1)}{R^2} \left( \xi_{lm}^{J+1} + \xi_{lm}^{J-1}
\right)  + \frac{3}{R} \frac{\xi_{lm}^{J+1} - \xi_{lm}^{J-1}}{2
\delta t} \\
&+&\frac{3}{4R^2} \left( \xi_{lm}^{J+1} + \xi_{lm}^{J-1}
\right) = \frac{l(l+1)}{2R^2} \left( \frac{\phi_{lm}^J(R)}{R} -
\left. \frac{\partial \phi_{lm}^J}{\partial r} \right|_{r=R} \right).
\end{eqnarray*}
This equation in $\xi_{lm}^{J+1}$ is solved and, 
for each pair $(l,m)$, we impose for~$\phi_{lm}^{J+1}$
$$
\left. \left( \frac{\partial}{\partial t}
+ \frac{\partial}{\partial r} + \frac{1}{r} \right) \phi_{lm}^{J+1}
(r) \right|_{r=R} = \xi_{lm}^{J+1} ,
$$
which looks like a modification of the condition (\ref{e:somerf}).

\subsection{Tests on outgoing waves}\label{ss:sorti}

The Sommerfeld boundary condition (\ref{e:somerf}) is an exact
condition, even at finite distance from the source, when only
considering monopolar waves. In order to test our implementation of
absorbing boundary condition (\ref{e:BC+}), we compared its efficiency
in being transparent to waves carrying only monopolar, dipolar and
quadrupolar terms, to the efficiency of the Sommerfeld boundary
condition for monopolar waves.  We started with $\phi=0$ at $t=0$ and
then solved Eq.~(\ref{e:defonde}) with
\begin{eqnarray}
\sigma(t,r,\theta,\varphi) &=& S(r,\theta,\varphi) e^{-1/t^2}
e^{-1/(t-1)^2} \qquad 0\leq t \leq 1 \label{e:defsig}\\
\sigma(t,r,\theta,\varphi) &=& 0 \text{ otherwise},\nonumber
\end{eqnarray}
with $S(r,\theta,\varphi)$ null for $r>R$.

In all cases, we performed a first calculation with a very large grid
(considered as infinite, we checked with various values of the radius
that the result in the interval $0\leq r\leq R$ would be the same), so
that in the time interval $[0,2R+1]$ the wave would not reach the
boundary, on which we imposed an homogeneous boundary
condition\footnote{results obtained here did not depend on the nature
of boundary conditions}. This gave us the reference solution crossing
the $r=R$ sphere without any reflection. We then solved again the same
problem, but on a grid of radius $R$, imposing Sommerfeld boundary
conditions $B_1\phi=0$ (\ref{e:somerf}), or our quadrupolar boundary
conditions $B_3\phi=0$ through the system
(\ref{e:defxi})-(\ref{e:ondsph}). The $L_1$ norm of the relative
difference between the functions obtained on the small grid and the
reference solution was taken as the error.

\begin{figure}
\centerline{\includegraphics[height=8cm]{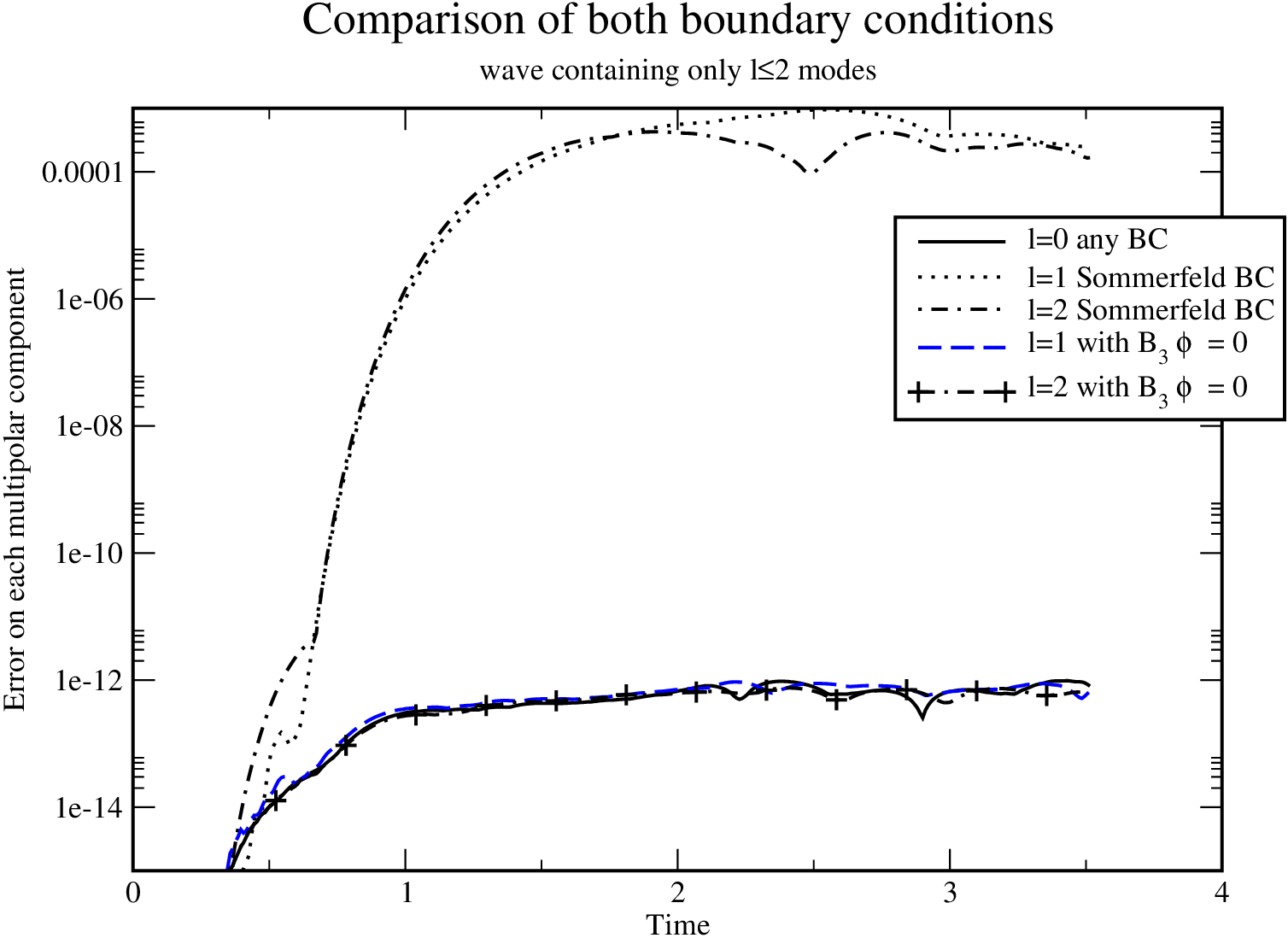}}
\vskip12pt
\caption{Comparison between the efficiencies of $B_1\phi=0$
  (\ref{e:somerf}) and $B_3\phi=0$ (\ref{e:defxi}) for $l\leq2$
  modes. The source of the wave equation is defined in
  Eqs.~(\ref{e:defsig}) and (\ref{e:defSl2}). We took $R=1.2$, a
  time-step $\delta t = 10^{-6}$, 33 polynomials for radial
  decomposition, 5 for $\theta$ and 4 for $\varphi$.  }
 \label{f:quadrup}
\end{figure}

\subsubsection{$l \leq 2$ case}
First, we took 
\begin{equation}
S(r,\theta,\varphi) = \left(e^{-r^2} - e^{-R^2} \right) \left(
r^2\cos^2\theta + r\sin \theta \cos \varphi \right),
\label{e:defSl2} 
\end{equation}
which contains only $l\leq 2$ modes. Figure~\ref{f:quadrup} shows the
relative efficiency of $B_3\phi=0$ (\ref{e:defxi}) condition to
$B_1\phi=0$ (\ref{e:somerf}) for all three modes present in the wave
generated by (\ref{e:defSl2}). For the monopolar ($l=0$) mode, the
evolution of the error would be the same for both types of boundary
conditions, within one percent of difference on the error. As far as
the discrepancy for dipolar and quadrupolar modes is concerned, one
can see that it drops from $10^{-4}$ with Sommerfeld boundary
condition, to $10^{-12}$ with $B_3\phi=0$ (\ref{e:defxi}). This
lower level is the same as for the monopolar mode with the Sommerfeld
boundary condition. We have checked that all solutions had converged
with respect to the number of spectral coefficients and to the
time-step. The error level at $10^{-12}$ is then mainly due to the
condition number of the matrix operator we invert (see
Sec.~\ref{ss:specmeu} above). We here conclude that our formulation of
$B_3\phi=0$ (\ref{e:defxi}) is as efficient for waves containing only
$l\leq 2$ modes as the Sommerfeld boundary condition (\ref{e:somerf})
for monopolar waves.

\begin{figure}
\centerline{\includegraphics[height=7cm]{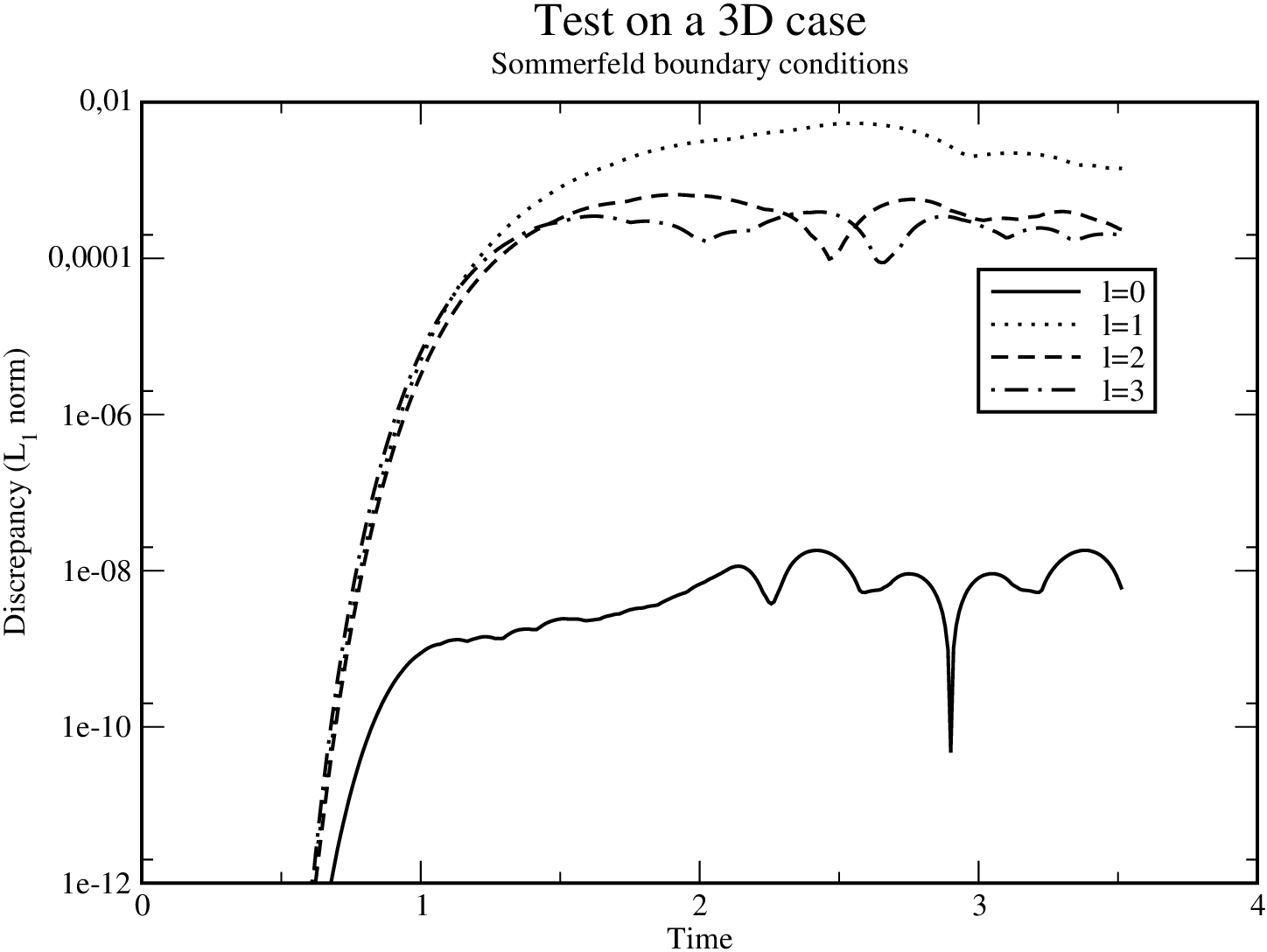}}
\vskip12pt
\caption{Time evolution of the first four modes of the wave generated
  by the source defined in Eqs.~(\ref{e:defsig}) and (\ref{e:defS3d});
  using $B_1\phi=0$ (\ref{e:somerf}). We took $R=1.2$, a time-step
  $\delta t = 10^{-4}$, 33 polynomials for radial decomposition, 17
  for $\theta$ and 16 for $\varphi$.  }
 \label{f:som3d}
\end{figure}
\begin{figure}
\centerline{\includegraphics[height=7cm]{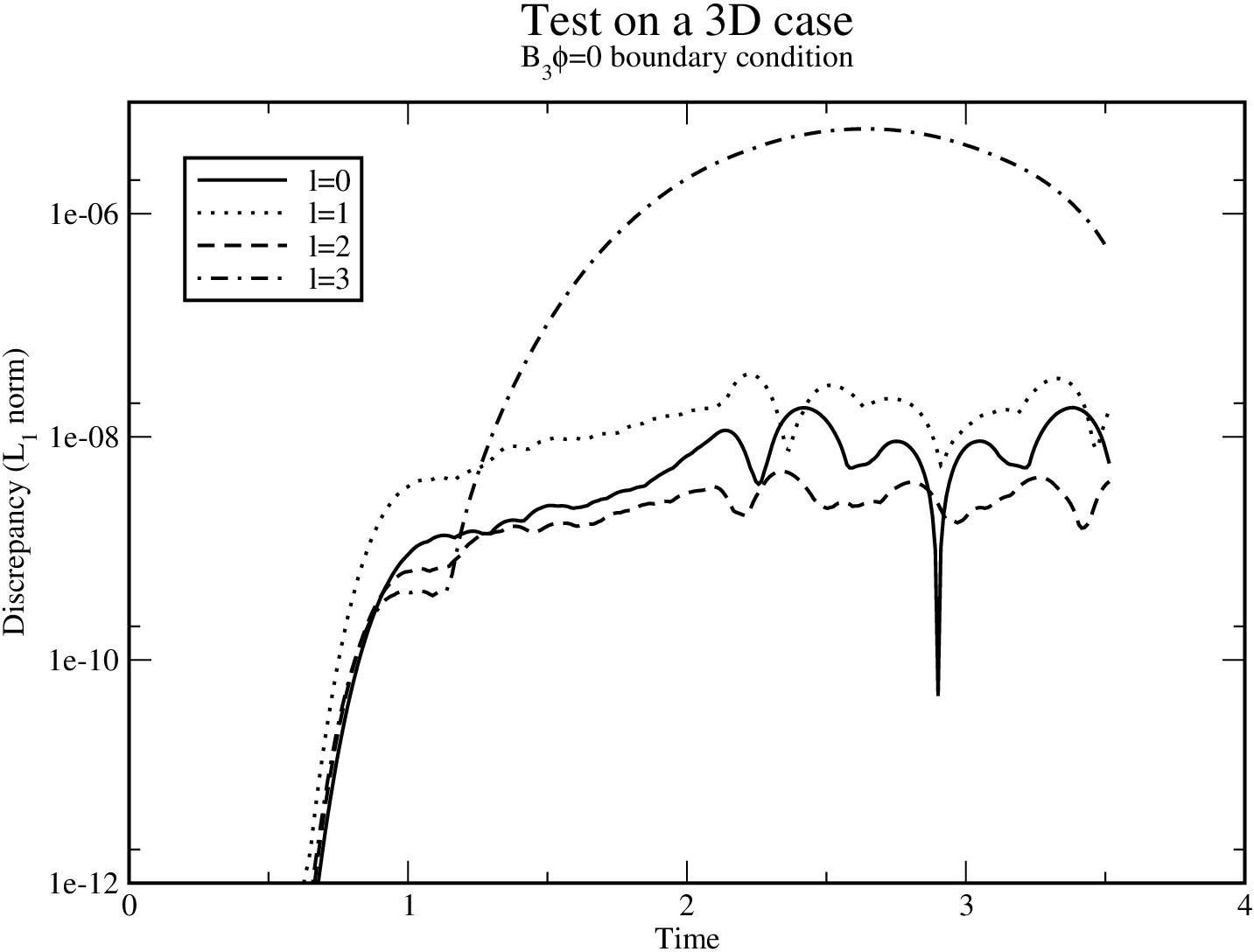}}
\vskip12pt
\caption{Time evolution of the first four modes of the wave generated
  by the source defined in Eqs.~(\ref{e:defsig}) and (\ref{e:defS3d});
  using $B_3 \phi = 0$ (\ref{e:defxi}) as the boundary condition. We
  took $R=1.2$, a time-step $\delta t = 10^{-4}$, 33 polynomials for
  radial decomposition, 17 for $\theta$ and 16 for $\varphi$.  }
 \label{f:bc33d}
\end{figure}
\subsubsection{Waves containing higher multipoles}
The study has been extended to a more general source $\sigma$
which contains {\it a priori} all multipolar terms:
\begin{equation}
S(r,\theta,\varphi) = \left(e^{-r^2} - e^{-R^2} \right)
\left(e^{-4(x-0.7)^2} + e^{-3(x+0.5)^2} \right). \label{e:defS3d} 
\end{equation}
Of course, in numerical
implementation, only a finite number of these terms are represented. 
The geometry of this source can be related to the distribution of mass
in the case of a binary system of gravitating bodies, which is one of
the main astrophysical sources of gravitational radiation we try to model. 
Let us make a comparison between the errors obtained,
on the one hand with the condition $B_1\phi=0$
(Figure~\ref{f:som3d}), and on the other hand with 
$B_3\phi=0$ (Figure~\ref{f:bc33d}).

\begin{figure}
\centerline{\includegraphics[height=7cm]{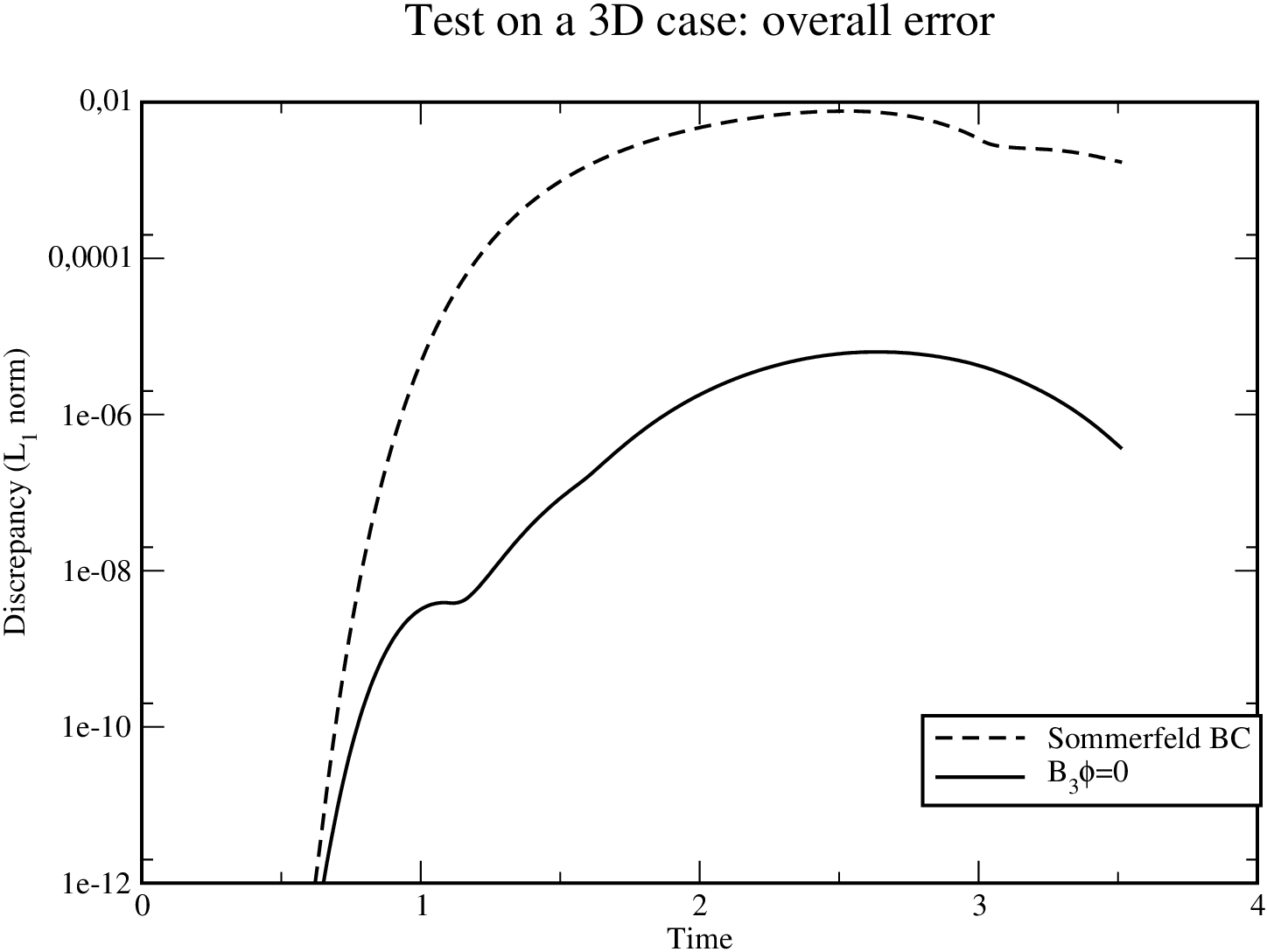}}
\vskip12pt
\caption{Time evolution of the error made in the computation of the
  wave generated by the source defined in Eqs.~(\ref{e:defsig}) and
  (\ref{e:defS3d}); using $B_1\phi=0$ (\ref{e:somerf}) and $B_3 \phi =
  0$ (\ref{e:defxi}). We took $R=1.2$, a time-step $\delta t =
  10^{-4}$, 33 polynomials for radial decomposition, 17 for $\theta$
  and 16 for $\varphi$.  }
 \label{f:err3d}
\end{figure}

As in the case in Figure~\ref{f:quadrup}, the error in the monopolar
component remains roughly the same, regardless of whether one uses
boundary condition (\ref{e:somerf}) or (\ref{e:defxi}). The errors for
the dipolar and quadrupolar components also exhibit similar
properties: the use of condition (\ref{e:defxi}) causes these errors
to be of the same magnitude as the error in the monopolar term. In the
case of Figure~\ref{f:bc33d}, this level is higher than on
Figure~\ref{f:quadrup} because a longer time-step has been
used. Finally, we have also plotted the discrepancies between the
reference and test solutions for the $l=3$ multipole. Following
\cite{BayT80}, the boundary condition $B_3 \phi = 0$ is not exact for
this component. Nevertheless, one can see a reduction in the error for
this component. This can be understood using the result of
\cite{BayT80} which shows that the condition $B_3 \phi = 0$ cancels
the first 3 terms in the asymptotic development in powers of $1/r$ of
the solution $\phi$ (\ref{e:asymp}). Then, since a given multipolar
term $l_0$ is present in terms like $1/r^n$ with $n\leq l_0$ (see
e.g.~\cite{Som49}), it is clear that the condition $B_3\phi = 0$ is
supposed to cancel all terms decaying slower than $1/r^4$ in the
$l\geq 3$ mode. Thus, the error displayed on Figure~\ref{f:err3d} is
three orders of magnitude lower with the condition $B_3\phi=0$ than
with $B_1\phi=0$.

We have checked this point, namely that the maximal error over the time
interval would decrease like $1/R^4$, where $R$ is the distance at which
the boundary conditions were imposed. We have also checked that the error
decreased both exponentially with the number of coefficients used in $r,
\theta$ or $\varphi$, as one would expect for spectral methods, and
like $\delta t^2$ (second-order time integration scheme).
Figure~\ref{f:err3d} shows the overall error as a function of time
for both boundary conditions used. Comparing Figure~\ref{f:err3d} with
figures~\ref{f:som3d} and \ref{f:bc33d}, one can see that most of the
error comes from the $l=1$ term when using $B_1\phi=0$ boundary
condition, and from the $l=3$ term when using $B_3\phi =0$. 
Finally, the computational cost of this enhanced boundary
condition is very low with this new approach. For the tests
presented here, the difference in CPU time would be of about
10\%. This is linked with the fact that our formulation
(\ref{e:defxi}) is a perturbation of the Sommerfeld boundary condition
(\ref{e:somerf}), where the quantity $\xi_{lm}(t)$ is obtained by
simple (ordinary differential equation) integration.

\section{Conclusions}\label{s:conc}

The purpose of this paper has been to provide a boundary condition that
is well-adapted for the simulation of astrophysical sources of
gravitational radiation, whose dominant modes are quadrupolar. We
took the series of boundary conditions derived by Bayliss and Turkel
\cite{BayT80}, truncated at quadrupolar order, and derived a new
formulation of that third-order condition in terms of a first-order
condition (resembling the classical radiation one), combined with a
wave-like equation on the outer boundary of the integration domain. This
formulation is simple in the sense that mixed derivatives are (almost)
absent. 

The numerical implementation using spectral methods and spherical
coordinates is straightforward and this formulation of high-order
boundary conditions requires only
a little more CPU time (less than 10\% in our tests) than the simplest
first-order condition (\ref{e:somerf}). We have verified that our
implementation of 
this boundary condition had the same efficiency with respect to transparency
for dipolar and quadrupolar waves as the Sommerfeld condition
(\ref{e:somerf}) for monopolar waves. The precision increases very
rapidly (like $1/R^4$) as one imposes the boundary condition further from
the source of radiation. These two points are of great interest for the
simulation of gravitational radiation from isolated astrophysical
sources. 

As an alternative, one can cite that more
accurate results may be obtained using the so-called 2+2 formalism in
the wave zone \cite{dIn97} and matching it to the results in 3+1
formalism\footnote{this is the {\em characteristic-Cauchy} matching,
CCM} near the source. Our approach is different, much simpler to implement
and should give accurate enough results for the Einstein equations.



\begin{thebibliography}{99}

\bibitem{cm1}
J. Novak, 
Review of Numerical Relativity Session
in 
\textit{Proc. of the ninth Marcel Grossman Meeting on General
Relativity, Rome, Italy, July 2000,}
edited by Jantzen, Gurzadyan and Ruffini, 
(World Scientific, Singapore, 2002). 

\bibitem{Yor79}
J.W. York,
Kinematics and dynamics in general relativity
in
\textit{Sources of gravitational radiation,}
edited by L. Smarr
(Cambridge University Press, Cambridge, 1979). 

\bibitem{SY78}
L. Smarr and J.W. York,
Radiation gauge in general relativity,
\textit{Phys. Rev. D}
\textbf{17},
1945 (1978).

\bibitem{Phil}
P. Grandcl\'ement, S.Bonazzola, E. Gourgoulhon and J.-A. Marck,
A multidomain spectral method for scalar and vectorial Poisson
equations with noncompact sources,
\textit{J. Comput. Phys. }
\textbf{170}, 
231 (2001), doi:10.1006/jcph.2001.6734.

\bibitem{Nov98}
J. Novak,
Spherical neutron star collapse toward a black hole in a tensor-scalar
theory of gravity,
\textit{Phys. Rev. D}
\textbf{57},
4789 (1998).

\bibitem{FriG00}
S. Fritelli and R. Gomez,
Ill-posedness in the Einstein equations,
\textit{J. Math. Phys. }
\textbf{41},
5535 (2000).

\bibitem{Bla00}
L. Blanchet,
Gravitational radiation from post-Newtonian sources and inspiralling
compact binaries,  
\textit{Living Rev. in Relativity}
\textbf{5},
3 (2002).

\bibitem{Hes98}
J.S. Hesthaven,
On the Analysis and Construction of Perfectly Matched Layers for the
Linearized Euler Equations,
\textit{J. Comput. Phys. }
\textbf{142},
129 (1998).

\bibitem{LioMV02}
J.-L. Lions, J. M\'etral and O. Vacus,
Well-posed absorbing layer for hyperbolic problems,
\textit{Numer. Math. }
\textbf{92},
535 (2002), doi:10.1007/s002110100263.

\bibitem{BayT80}
A. Bayliss and E. Turkel,
Radiation boundary conditions for wave-like equations,
\textit{Comm. Pure and Appl. Math.}
\textbf{33},
707 (1980).

\bibitem{Som49}
A. Sommerfeld,
\textit{Partial Differential Equations in Physics},
(Academic Press, New York, 1949).

\bibitem{AlaDR02}
U.E. Aladl, A.S. Deakin and H. Rasmussen,
Nonreflecting boundary condition for the wave equation,
\textit{J. Comput. Appl. Math.}
\textbf{138},
309 (2002).

\bibitem{EngM77}
B. Engquist and A. Majda,
Absorbing boundary conditions for the numerical simulation of waves,
\textit{Mathematics of Computation}
\textbf{31},
629 (1977).

\bibitem{GotO77}
D. Gottlieb and S.A. Orszag,
\textit{Numerical Analysis of Spectral Methods: Theory and Applications},
(Society for industrial and applied mathematics, Philadelphia, 1977). 

\bibitem{CanHQZ88}
C. Canuto, M.Y. Hussaini, A. Quarteroni and T.A. Zhang, 
\textit{Spectral Methods in Fluid Dynamics},
(Springer-Verlag, Berlin, 1988). 

\bibitem{Boy01}
J.P. Boyd,
\textit{Chebyshev and Fourier Spectral Methods},
(Dover Publication, Mineola, 2001).

\bibitem{BonGM99}
S. Bonazzola, E. Gourgoulhon and J.-A. Marck,
Spectral methods in general relativistic astrophysics,
\textit{J. Comput. Appl. Math.}
\textbf{109},
433 (1999).

\bibitem{dIn97}
R.A. d'Inverno, 
2+2 formalism and applications
in 
\textit{Relativistic Gravitation and Gravitational Radiation,}
edited by J.-A. Marck and J.-P. Lasota, 
(Cambridge University Press, Cambridge, 1997). 

\end{thebibliography}
\end{document}